\documentclass{emulateapj}
\usepackage{apjfonts}

\lefthead{HAN} \righthead{MICROLENSING MULTIPLE-PLANET ANALYSIS}

\newcommand{\zvec}{\mbox{\bf z}}

\newcommand{\thetae}{\theta_{\rm E}}

\begin{document}
\title{Analysis of Microlensing Light Curves Induced by Multiple-Planet Systems}

\author {Cheongho Han}
\affil{Department of Physics, Institute for Basic Science
Research, Chungbuk National University, Chongju 361-763, Korea,
cheongho@astroph.chungbuk.ac.kr}



\begin{abstract}
To maximize the number of planet detections by increasing efficiency, 
current microlensing follow-up observation experiments are focusing on 
high-magnification events to search for planet-induced perturbations 
near the peak of lensing light curves.  It was known that by monitoring 
high-magnification events, it is possible to detect multiplicity signatures 
of planetary systems.  However, it was believed that the interpretation 
of the signals and the characterization of the detected multiple-planet 
systems would be difficult due to the complexity of the magnification 
pattern in the central region combined with the large number of lensing 
parameters  required to model multiple-planet systems.  In this paper, 
we demonstrate that in many cases the central planetary perturbations 
induced by multiple planets can be well approximated by the superposition 
of the single planetary perturbations where the individual planet-primary 
pairs act as independent binary lens systems (binary superposition).  
The validity of the binary-superposition approximation implies that the 
analysis of perturbations induced by multiple planets can be greatly 
simplified because the anomalies produced by the individual planet 
components can be investigated separately by using relatively much 
simpler single-planetary analysis, and thus enables better characterization 
of these systems. 
\end{abstract}


\section{Introduction}

Microlensing is one of the most important methods that can detect and 
characterize extrasolar planets (see the review of Perryman 2000).  
Microlensing planet detection is possible because planets can induce
perturbations to the standard lensing light curves produced by primary 
stars.  Once the perturbation is detected and analyzed, one can determine 
the mass ratio, $q$, and the projected separation, $s$ (normalized by the 
Einstein ring radius $r_{\rm E}$), between the planet and host star 
\citep{mao91}.  Recently, a clear-cut microlensing detection of an 
exoplanet was reported by \citet{bond04}.

Planet detection via microlensing is observationally challenging.  One of 
the most important difficulties in detecting planets via microlensing lies 
in the fact that planet-induced perturbations last for a short period of 
time.  For a Jupiter-mass planet, the duration is only a few days and it 
decreases as $\propto \sqrt{q}$.  To achieve high monitoring frequency 
required for planet detections, current lensing experiments are employing 
early warning system to issue alerts of ongoing events in the early stage 
of lensing magnification \citep{soszynski01, bond01} and follow-up observation 
programs to intensively monitor the alerted events \citep{bond02, park04, 
cassan04}.  However, follow-up is generally done with small field-of-view 
instrument, and thus events should be monitored sequentially.  As a result, 
only a handful number of events can be followed at any given time, limiting 
the number of planet detections.

An observational strategy that can dramatically increase the planet 
detection efficiency was proposed by \citet{griest98}.  When a microlensing 
event is caused by a star possessing a planet, two sets of caustics are 
produced.  Among them, one is located away from the primary lens (planetary 
caustic) and the other is located close to the primary lens (central caustic). 
The location of the planetary caustic varies depending on the planetary 
separation $s$, which is not known, and thus it is impossible to predict 
the time of planetary perturbation in advance.  On the other hand, the 
central caustic is always located very close to the primary lens, and thus 
the central perturbation occurs near the peak of high-magnification events.  
Therefore, by focusing on high-magnification events, it is possible to 
dramatically increase the planet detection efficiency, enabling one to 
maximize the number of planet detections with a limited use of resources 
and time \citep{han01b, rattenbury02, yoo04}.

An additional use of high magnification events was noticed by \citet{gaudi98}.  
They pointed out that multiple planets with separations $\sim 0.6$ -- $1.6$ 
of the Einstein ring radius $r_{\rm E}$ significantly affect the central 
region of the magnification pattern regardless of the orientation and thus 
microlensing can be used to detect multiple-planet systems.  They noted, 
however, that characterizing the detected multiple-planet systems by 
analyzing the central perturbations would be difficult due to the complexity 
of the magnification pattern combined with the large number of lensing 
parameters required to model multiple-planet systems.

\begin{figure*}[htb]
\epsscale{0.85}
\caption{\label{fig:one}
Contour maps of magnification excess from the single-mass lensing as a 
function of the source position $(\xi,\eta)$ for a lens system containing 
two planets.  The parameters of planet 1 (located on the left-side $\xi$ 
axis) are held fixed at $q_1=0.001$, $s_1=1.4$, while the projected 
separation $s_2$ and the angle between the axes, $\Delta\theta$, are 
varied for a second planet with $q_2=0.0003$.  Greyscale is used to 
represent positive (bright) and negative (dark) deviation regions.  The 
two sets of contours drawn in black and white are for the maps constructed 
by exact triple lensing and binary-superposition approximation, respectively.  
Contours are drawn at the levels of $|\epsilon|=2$\%, 5\%, and 10\%.  The 
straight line with an arrow in each panel is the source trajectory where 
the resulting light curve is presented in the corresponding panel of 
Fig.~\ref{fig:four}.  Lengths are scaled by the Einstein ring radius 
$\theta_{\rm E}$.  The coordinates are centered at the effective position 
of the primary lens [see eq.~(\label{eq7}) for definition].  The orientation 
is such that $\Delta\theta=0^\circ$ implies the two planets are on the same 
side.
}\end{figure*}

In this paper, we demonstrate that in many cases the central planetary 
perturbations induced by multiple planets can be well approximated by the 
superposition of the single-planetary perturbations where the individual 
planet-primary pairs act as independent binary lens systems (binary 
superposition).  The validity of the binary-superposition approximation 
implies that a simple single-planet lensing model is possible for the 
description of the anomalies produced by the individual planet components, 
enabling better characterization of these systems.

The layout of the paper is as follows.  In \S\ 2, we describe the 
multiple-planetary lensing and magnification pattern in the central region.  
In \S\ 3, we illustrate the validity of th binary-superposition approximation 
in describing the central magnification pattern of multiple-planet systems.  
In \S\ 4, we discuss the usefulness of the superposition approximation in 
the interpretation of the multiple planetary signals.

\begin{figure*}[htb]
\epsscale{0.85}
\caption{\label{fig:two}
Enlargement of the excess magnification maps presented in Fig.~\ref{fig:one}.  
The figures drawn in solid black and white lines are the caustics for the 
cases of the exact triple lensing and binary-superposition approximation, 
respectively.  Contours and greyscale are drawn by the same way as in 
Fig.~\ref{fig:one}.
}\end{figure*}

\section{Multiple Planetary Lensing}

The equation of lens mapping from the lens plane to the source plane (lens 
equation) of an $N$ point masses is expressed as
\begin{equation}
\zeta = z - \sum_{j=1}^N {m_j/M \over \bar{z}-\bar{z}_{L,j}},
\label{eq1}
\end{equation}
where $\zeta=\xi + i\eta$, $z_{L,j}=x_{L,j}+iy_{L,j}$, and $z=x+iy$ are the 
complex notations of the source, lens, and image positions, respectively, 
$\bar{z}$ denotes the complex conjugate of $z$, $m_j$ are the masses of 
the individual lens components, $M=\sum_j m_j$ is the total mass of the 
system, and thus $m_j/M$ represent the mass fractions of the individual 
lens components.  Here all angles are normalized to the Einstein ring 
radius of the total mass of the system, $M$, i.e.
\begin{equation}
\thetae={r_{\rm E} \over D_{\rm OL}}=\left[ {4GM\over c^2} 
\left( {1\over D_{\rm OL}} - {1\over D_{\rm OS}}  \right)
\right]^{1/2},
\label{eq2}
\end{equation}  
where $D_{\rm OL}$ and $D_{\rm OS}$ are the distances to the lens and 
source, respectively.  The lensing process conserves the source surface 
brightness, and thus the magnifications $A_i$ of the individual images 
correspond to the ratios between the areas of the images and source.  
For an infinitesimally small source element, the magnification is,
\begin{equation}
A_i = \left\vert \left( 1-{\partial\zeta\over\partial\bar{z}}
{\overline{\partial\zeta}\over\partial\bar{z}} \right)^{-1} \right\vert.
\label{eq3}
\end{equation}
The total magnification is the sum over all images, $A=\sum_i A_i$.

For a single lens ($N=1$), the lens equation is simply inverted to solve 
the image positions $(x,y)$ and magnifications for given lens $(x_L,y_L)$ 
and source $(\xi,\eta)$ positions.  This yields the familiar result that 
there are two images with magnifications and separations from the lens of 
$A_\pm=0.5(A \pm 1)$ and $u_{I,\pm}=|z_\pm-z_L|= 0.5[u \pm (u^2 +4)^{1/2}]$, 
respectively, where $u\equiv |\zeta-z_{L}|$ is the separation between the 
lens and source and $A=A_{+} + A_{-}=(u^2+2)/[u(u^2+4)^{1/2}]$ is the 
total magnification.

A planetary lensing with a single planet is described by the formalism of 
a binary ($N=2$) lens.  In this case, the lens equation cannot be inverted 
algebraically.  However, it can be expressed as a fifth-order polynomial 
in $z$ and the image positions are then obtained by numerically solving 
the polynomial \citep{witt95}.  One important characteristic of binary 
lensing is the formation of caustics, which represent the set of source 
positions at which the magnification of a point source becomes infinite. 
Planetary perturbations on lensing light curves occur when the source 
approaches close to the caustics.  The location and size of these 
caustics depend on the projected separation $s$ and the mass ratio $q$.  
For a planetary case, there exist two sets of disconnected caustics.  
The planetary caustic(s) is (are) located away from the primary star on 
or very close to the star-planet axis with a separation from the primary 
star of $\xi_{\rm pc} \simeq s-1/s$.  The central caustic, on the other 
hand, is located close to the primary lens with a size of 
$\Delta \xi_{\rm cc} \simeq 4q/(s-1/s)^2$ \citep{chung05}.  Caustics are 
located within the Einstein ring when the planetary separation is within 
the range of $0.6 \lesssim s \lesssim 1.6$.  The size of the caustic, 
which is directly proportional to the planet detection efficiency, is 
maximized when the planet is located within this range, and thus this 
range is referred as ``lensing zone'' 
\citep{gould92}.

For a multiple-lens system, the lens equation is equivalent to a polynomial 
with an order of ($N^2+1)$.  Therefore, a multiple-lens system produces a 
maximum of $N^2+1$ and a minimum of $N+1$ images, with the number of images 
changing by multiple of two as the source crosses a caustic \citep{rhie97}.  
For a system with two planets (and thus $N=3$), there are thus a maximum of 
10 images and a minimum of 4 images.  Unlike the caustics of binary lensing, 
those of multiple lensing can exhibit self-intersecting and nesting.

\begin{figure*}[htb]
\epsscale{0.85}
\plotone{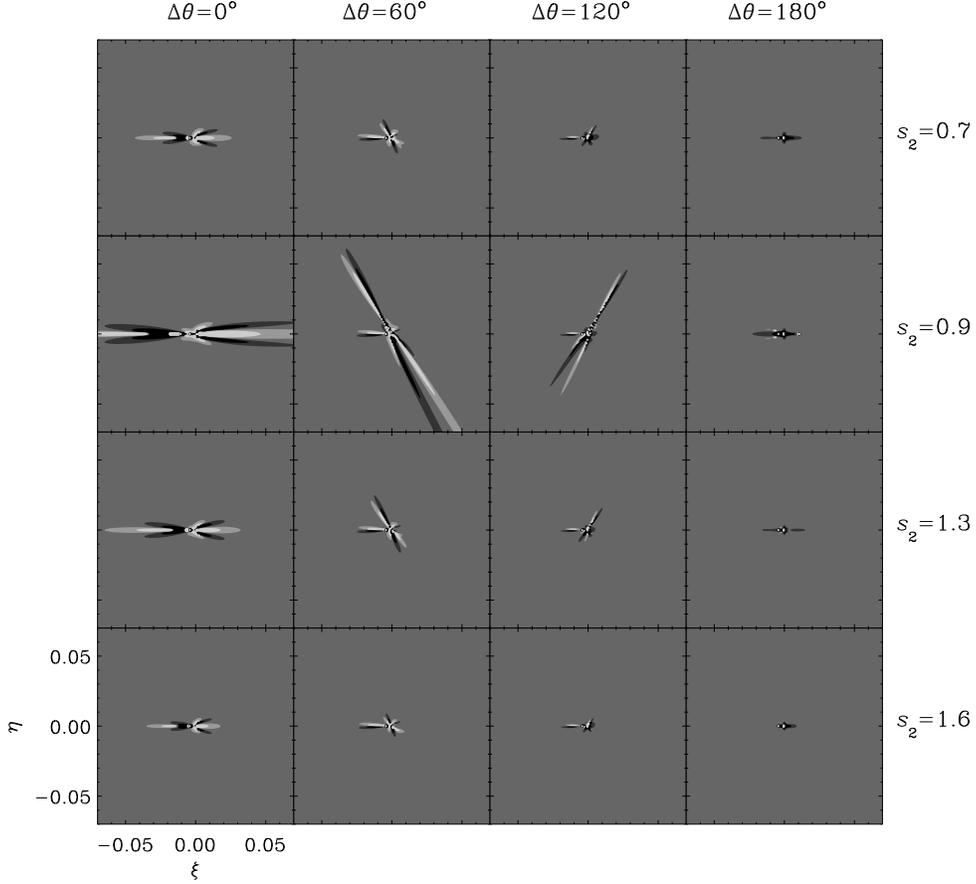}
\caption{\label{fig:three}
Greyscale maps of the difference between the magnification excesses 
of the triple lensing and binary approximation.  i.e.\ $\Delta\epsilon=
\epsilon_{\rm bs}-\epsilon_{\rm tri}=(A_{\rm bs}-A_{\rm tri})/A_0$.
Greyscales brighter and darker than background represent the region 
where $A_{\rm tri}>A_{\rm bs}$ and vice versa, respectively, and the 
scale changes at $|\epsilon_{\rm bs}|=2\%$ and $5\%$.
}\end{figure*}

\begin{figure*}[htb]
\epsscale{0.85}
\caption{\label{fig:four}
Example microlensing light curves produced by multiple-planet systems. 
The parameters of the planets are identical to those in Fig.~\ref{fig:one}.
The source trajectories responsible for the events are marked also in 
Fig.~\ref{fig:one}.  For all events, the impact parameter of the 
trajectories is $u_0=0.03$.  In each panel, there are five curves, 
where the curves drawn in thick solid black, thin magenta, cyan, and red 
lines are those of exact triple lensing, binary lensing with the pairs 
of the primary and the individual planets, and binary-superposition 
approximation, respectively, and the dashed curve is that of a single-mass 
lensing of the primary alone.  The curve in the lower part of each panel 
shows the difference between the excesses of the exact triple lensing 
and binary-superposition approximation.
}\end{figure*}

\section{Binary-Superposition Approximation}

The mass ratio of a planet to its primary star is very small, and thus 
the lensing behavior of planet-induced anomalies can be treated as 
perturbation \citep{gould92, bozza99, asada02, dominik99, an05}.  Due 
to the perturbative nature of planetary anomalies, it was known that 
the magnification pattern in the region around {\it planetary} caustics 
of multiple-planet systems can be described by the superposition of those 
of the single-planet systems where the individual planet-primary pairs 
acts as independent binary lens systems \citep{han01a}, i.e.,
\begin{equation}
A_{\rm mul}\sim A_{\rm bs}=\sum_{i=1}^{N-1} A_{{\rm bi},i} - A_0,
\label{eq4}
\end{equation}
where $A_{\rm mul}$, $A_{\rm bs}$, $A_{{\rm bi},i}$, and $A_0$ represent 
the magnifications of the exact multiple lensing, binary-superposition 
approximation, binary lensing with the pairs of the primary and individual 
planets, and single-mass lensing of the primary alone, respectively.
By contrast, it was believed that the binary-superposition approximation 
would not be adequate to describe the magnification pattern in the central 
region because the central caustics produced by the individual planet 
components reside at the same central region and thus non-linear 
interference of the perturbations would be large.

Unlike this belief about the magnification pattern in the central 
perturbation region, we find that non-linear interference between the 
perturbations produced by the individual planets of a multiple-planet 
system is important only in a small confined region very close to the 
central caustics.  This implies that in many cases binary-superposition 
approximation can also be used for the description of the magnification 
pattern in the central perturbation region.

To demonstrate the validity of the binary-superposition approximation 
in the central region, we construct a set of {\it magnification excess} 
maps of example multiple-planet systems containing two planets with 
various orientations.  The magnification excess represents the deviation 
of the magnification from the single-mass lensing as a function of the 
source position $(\xi,\eta)$, and it is computed by
\begin{equation}
\epsilon_{\rm tri}={A_{\rm tri}-A_{0} \over A_{\rm 0}},
\label{eq5}
\end{equation}
where $A_{\rm tri}$ is the magnification of the triple (primary star 
plus two planets) lensing.

In Figure~\ref{fig:one}, we present the constructed contour (drawn by 
black lines) maps of magnification excess.  In the map, the parameters 
of planet 1 are held fixed at $q_1=0.001$ and $s_1=1.4$, while the 
projected separation $s_2$ and the angle between the position vectors 
to the individual planets from the primary star (orientation angle), 
$\Delta\theta$, are varied for a second planet with $q_2=0.0003$.  
Greyscale is used to represent positive (bright, $\epsilon_{\rm tri}>0$) 
and negative (dark) deviation regions.  Also drawn are the contours 
(drawn in white lines) of magnification excess based on binary-superposition 
approximation, i.e., 
\begin{equation}
\epsilon_{\rm bs}=\sum _{i=1}^{2} {A_{{\rm bi},i}-A_{0} \over A_{\rm 0}}.
\label{eq6}
\end{equation}
For the maps based on binary superposition, we consider slight shift of 
the effective lensing position of the primary star ($\zvec_{L,\ast}$) 
toward the individual planets ($\zvec_{L,p_{i}}$) with an amount of
\citep{distefano96}
\begin{equation}
\delta \zvec_{L,\ast} = \sum_{i=1}^2 
{{q_i}\over s_i+1/s_i} {\zvec_{L,p_{i}} - \zvec_{L,\ast}\over 
|\zvec_{L,p_{i}} - \zvec_{L,\ast}| }.
\label{eq7}
\end{equation}
To better show the magnification pattern in the very central region and 
the detailed caustic structure, we enlarge the maps and presented them 
in Figure~\ref{fig:two}.  In each map, the figures drawn in thick black 
and white lines represent the caustics for the cases of the exact triple 
lensing and binary superposition, respectively.

From the comparison of the excess maps constructed by the exact triple 
lensing and binary-superposition approximation, one finds that binary 
superposition is a good approximation in most of the central perturbation 
region as demonstrated by the good match between the two sets of contours.  
Slight deviation of the binary-superposition approximation from the exact 
lensing magnification occurs (a) in a small region very close to the 
central caustics and (b) in the narrow regions along the primary-planet 
axes.  This can be better seen in Figure~\ref{fig:three}, where we present 
the greyscale maps for the difference between the excesses of the triple 
lensing and binary-superposition approximation, i.e., 
\begin{equation}
\Delta\epsilon = \epsilon_{\rm tri}-\epsilon_{\rm bs}.
\label{eq8}
\end{equation}
The difference in the region close to the central caustics is caused by 
the non-linear interference between the perturbations produced by the 
individual planets.  On the other hand, the difference along the 
primary-planet axes is caused by the slight positional shift of the 
triple-lensing caustics due to the introduction of an additional planet.  
However, the area of the deviation region, in general, is much smaller 
than the total area of the perturbation region, and thus the binary 
superposition approximation is able to well describe most part of 
planetary anomalies in lensing light curves.  This can be seen in 
Figure~\ref{fig:four}, where we present example light curves of 
multiple-planetary lensing events and compare them to those obtained by 
binary superposition.

\begin{figure}[htb]
\epsscale{1.1}
\plotone{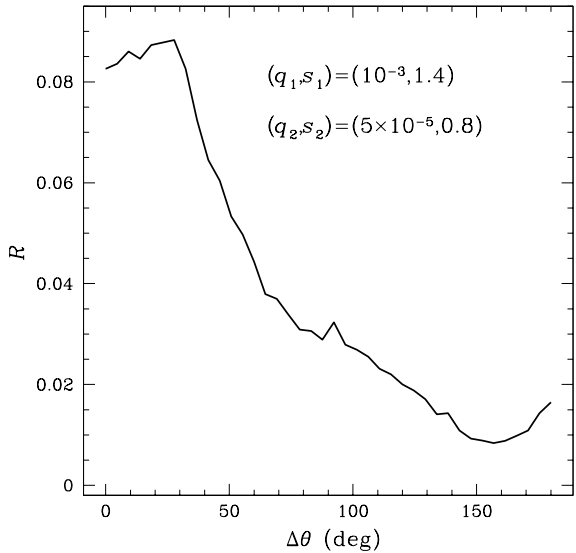}
\caption{\label{fig:five}
Dependence of the binary-superposition validity (${\cal R}$) on the 
angle  between the position vectors to the component planets from the 
host star ($\Delta\theta$).  See eq.~(\ref{eq9}) for the definition of 
${\cal R}$.  To see the dependence only on $\Delta\theta$, we fix 
$(q_1,s_1)$ and $(q_2,s_2)$ and their values are marked in the panel.
}\end{figure}

\begin{figure}[htb]
\epsscale{1.1}
\plotone{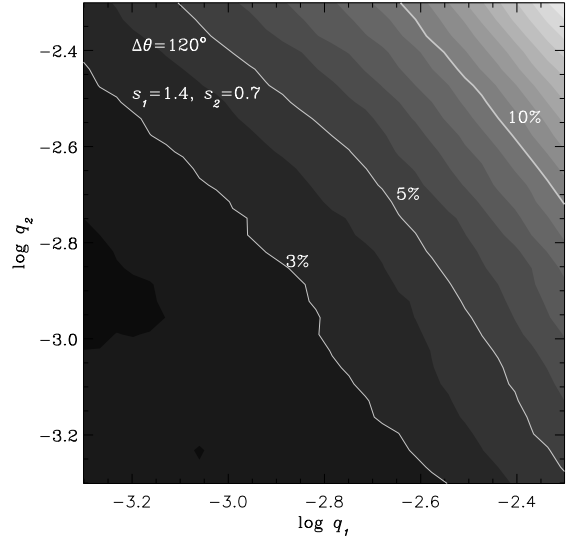}
\caption{\label{fig:six}
A contour map of the binary-superposition validity (${\cal R}$) as a 
function of the planet mass ratios of a multiple planetary system 
composed of two planets.  To see the dependence only on the mass ratios, 
we fix $s_1$, $s_2$ and $\Delta\theta$ and their values are marked in 
the panel.
}\end{figure}

The lensing behavior of a multiple planetary system is determined by many 
parameters including $q_1$, $q_2$, $s_1$, $s_2$, and $\Delta\theta$, and 
binary-superposition might not be a good approximation in some space region 
of these parameters.  We therefore investigate the region of parameter 
space where binary-superposition is a poor approximation.  Due to the 
numerousness of the parameters, we choose a method of investigation where 
we inquire the validity of the approximation on the individual parameters 
by varying one parameter and fixing other parameters.  The validity of the 
approximation is quantified by the ratio of
\begin{equation}
{\cal R}={\Sigma_{\Delta\epsilon} \over \Sigma_{\epsilon}},
\label{eq9}
\end{equation}
where $\Sigma_{\epsilon}$ and $\Sigma_{\Delta\epsilon}$ represent the area 
of the central perturbation region and the area of the difference region 
between the exact triple lensing and the binary-superposition approximation, 
respectively.  In some cases, the perturbation regions caused by the 
planetary and central caustics are connected, making the boundary between 
the two regions ambiguous.  We thus define the central perturbation region 
as the region within the lens-source impact parameter of $u=0.1$ 
(corresponding to the region with magnifications $A\gtrsim 10$).  The 
areas $\Sigma_{\Delta\epsilon}$ and $\Sigma_{\epsilon}$ are computed by 
setting the threshold values of $|\Delta\epsilon_{th}|=5\%$ and 
$|\epsilon_{th}|=5\%$, respectively.

\begin{figure}[htb]
\epsscale{1.1}
\plotone{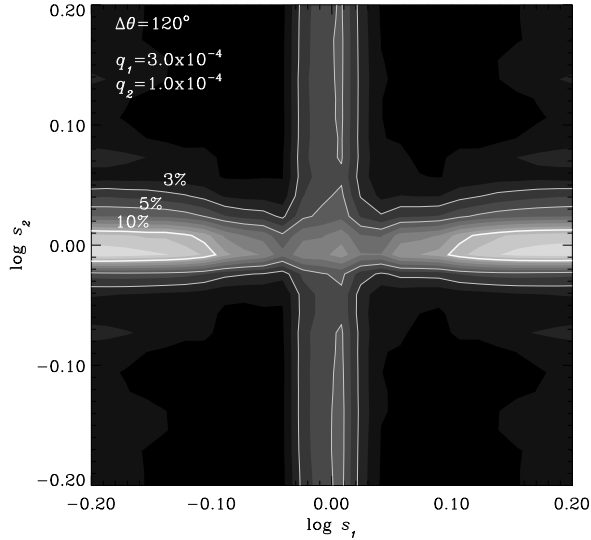}
\caption{\label{fig:seven}
A contour map of the binary-superposition validity (${\cal R}$) as a 
function of the separations to the planets from the host star of a 
multiple planetary system composed of two planets.  To see the dependence 
only on the separations, we fix $q_1$, $q_2$ and $\Delta\theta$ and their 
values are marked in the panel.
}\end{figure}

In Figures~\ref{fig:five}, \ref{fig:six}, and \ref{fig:seven}, we present 
the dependence of ${\cal R}$ on the orientation angle ($\Delta\theta$),
the mass ratios of the component planets ($q_1$ and $q_2$), and the 
separations to them ($s_1$ and $s_2$), respectively.  From the variation 
of ${\cal R}$ depending on the orientation angle, we find that the 
difference between the exact lensing and binary-superposition approximation 
becomes bigger as $\Delta\theta$ decreases and thus the two planets are 
located closer to each other.  We interpret this tendency as the increase 
of the non-linear interference between the perturbations caused by the two 
planets as the separation between them decreases.  From the dependence on 
the mass ratio and separation, we find that binary superposition becomes 
a poor approximation as either the planet mass increases or the separation 
approaches to unity.  These tendencies are the natural results of the 
breakdown of the perturbation treatment for companions with high mass 
ratios and (or) separations of $s\sim 1$, because the perturbation treatment 
is valid when $q\ll 1$ and $|s-1|\gg q$.  Besides these extreme regions of 
parameter space, however, we find that $R\lesssim 5\%$ in most regions, 
implying that binary-superposition approximation well describes the lensing 
behavior of most multiple planetary systems.

\section{Implication and Conclusion}

In the previous section, we demonstrated that lensing magnification 
patterns of multiple-planet systems can be described by using 
binary-superposition approximation not only in the region around planetary 
caustics but also in most part of the central perturbation region.  In 
this section, we discuss the importance of the binary-superposition 
approximation in the analysis and characterization of multiple-planet 
systems to be detected via microlensing.

Exact description of lensing behavior of events caused by multiple-planet 
systems requires a large number of parameters.  Even for the simplest case
of a two-planet system, the number of parameters is ten, including four 
single-lensing parameters of the Einstein timescale $t_{\rm E}$, time of 
the closest lens-source approach $t_0$, lens-source impact parameter $u_0$, 
and blended light fraction $f_B$, and another four parameters of the two 
sets of the planetary separation and mass ratio, $(s_1,q_1)$ and $(s_2,q_2)$, 
and the source trajectory angle $\alpha$ and the orientation angle 
$\Delta\theta$.  As a result, it was believed that analyzing anomalies 
produced by multiple planets would be a daunting task.

With the validity of binary-superposition approximation, however, the 
analysis can be greatly simplified because the anomalies induced by the 
individual planets can be investigated separately by using relatively 
much simpler single-planetary lensing analysis.  Anomalies for which this 
type of analysis is directly applicable are those where the perturbations 
induced by the individual planets are well separated, e.g., the anomalies 
in the lensing light curves presented in Figure~\ref{fig:four} with 
$(s_2,\Delta\theta)=(0.7, 60^\circ)$, $(0.7, 120^\circ)$, $(0.9, 60^\circ)$, 
$(0.9, 120^\circ)$, $(1.3, 60^\circ)$, and $(1.3, 120^\circ)$.

In some cases, the region of perturbations caused by the individual 
planets are located close together and thus part of the anomalies in 
lensing light curves can be blended together.  However, even in these 
cases, the non-linear interference between the anomalies is important 
only in small confined regions and thus the superposition approximation 
would still be valid for a large portion of the anomalies, allowing 
rough estimation of the separations and mass ratios of the individual 
planets.  Once these rough parameters are determined, then fine tuning of 
the parameters by using exact multiple-planet analysis will be possible 
by exploring the parameter space that was greatly narrowed down.

However, care is required for the case when the two perturbations caused
by the individual planets happen to locate at the same position (or very 
close to each other).  This case occurs when the two planets are aligned 
($\Delta\theta=180^\circ$) or anti-aligned ($\Delta\theta=0^\circ$).  In 
this case, the anomaly  appears to be caused by a single planet and thus 
naive analysis of the anomalies can result in wrong characterization of 
the planet system.  However, this type of anomalies will be very rare.

\acknowledgments 
We would like to thank A. Gould for making useful comments on the paper.
This work was supported by the Astrophysical Research Center for the 
Structure and Evolution of the Cosmos (ARCSEC") of Korea Science \& 
Engineering Foundation (KOSEF) through Science Research Program (SRC) 
program.

\end{document}